\documentstyle[11pt,fleqn]{article}
\newcommand{\R}{\mbox{$I\!\!R$}}             
\newcommand{\C}{\mbox{$I\!\!\!\!C$}}         

\begin{document}


\hfill{}
\par
\bigskip
\par
\rm


\par
\bigskip
\LARGE
\noindent
{\bf  Local $\zeta$-functions, stress-energy tensor, field fluctuations, and all that, in curved static spacetime}
\par
\bigskip
\par
\rm
\normalsize
\noindent {\em Dedicated to Prof. Emilio Elizalde on the occasion of his 60th birthday}
\par
\bigskip
\par
\rm


\large
\noindent {\bf Valter Moretti}

\large
\smallskip

\noindent
Department of Mathematics, Trento University,
I-38050 Povo (TN), Italy.\\
E-mail: moretti@science.unitn.it

\large
\smallskip

\rm\normalsize



\par
\bigskip
\par
\hfill{\sl October 2010}
\par
\medskip
\par\rm



\noindent
{\bf Abstract:}
This is a quick review on some technology concerning the local {\em zeta function} 
applied to Quantum Field Theory in curved static (thermal) spacetime to regularize  the stress energy tensor and the field fluctuations. 
\par

 \rm




\section{Quasifree QFT in curved static manifolds, Euclidean approach
$\zeta$-function technique.}

\noindent {\bf 1.1} {\em The $\zeta$-function determinant.} 
Suppose we are given  a $n\times n$ positive-definite Hermitian matrix  $A$ with 
eigenvalues $0<\lambda_1\leq \lambda_2\leq \cdots \leq \lambda_n$.
One can define the complex-valued function
\begin{eqnarray}
\zeta(s|A) = \sum_{j=1}^n \lambda_j^{-s}\:,
\end{eqnarray}
where $s\in \C$. (Notice that $\lambda_j^{-s}$ is well-defined since 
$\lambda_j>0$.)
By direct inspection one proves that:
\begin{eqnarray}
\det A = e^{-\frac{d\zeta(s|A)}{ds}|_{s=0}}\:. \label{det'}
\end{eqnarray}
This trivial result can be generalized to provide a useful definition of the determinant of an operator working in an infinite-dimensional Hilbert space.
To this end, focus on a non-negative self-adjoint operator $A$ whose spectrum is discrete and each eigenspace has a finite dimension, and  consider
 the series with $s\in \C$ (the prime on the 
sum henceforth  means that any possible null eigenvalues is omitted)
\begin{eqnarray}
\zeta(s|A) := {\sum_j}' \lambda^{-s}\:.
\label{sum0}
\end{eqnarray}
Looking at (\ref{det'}), the idea \cite{haw} is to define, once again.
\begin{eqnarray}
\det A = e^{-\frac{d\zeta(s|A)}{ds}|_{s=0}}\:, \nonumber
\end{eqnarray}
where now  the function $\zeta$ on the right-hand side is, in the general case,  the {\em analytic continuation
of the function defined by the series} (\ref{sum0}) in its convergence domain, since the series may diverge at $s=0$ -- and this is the 
standard situation in the infinite-dimensional case! -- provided that the analytic extension really reaches a neignorhood of the point
$s=0$. The interesting fact is that this procedure truly works in physically relevant cases, related to QFT in curved spacetime,
producing meaningful results as
 we go to discuss in the following section.\\

\noindent {\bf 1.2} {\em Thermal QFT in static spacetimes}. A smooth globally hyperbolic spacetime $(M,g)$  is said to be
{\em static} if  it admits a (local)  time-like Killing vector field $\partial_t$  normal to a smooth spacelike Cauchy surface 
$\Sigma$. Consequently, there are  (local) coordinate frames
$(x^0,x^1,x^2,x^3)\equiv (t,\vec{x})$ where $g_{0i} = 0$ ($i=1,2,3$) and $\partial_t g_{\mu\nu}=0$ and $\vec{x}$ are local coordinates on $\Sigma$.
Though the results we are going to present may be generalize to higher spin fields,  we henceforth stick to the case of a real scalar field $\phi$ propagating in $M$ and
satisfying an equation of motion of the form
\begin{eqnarray}
P \phi = 0 \label{evolution}\:,
\end{eqnarray}
where $P := -\nabla_\mu\nabla^\mu + V$, $V$ being a smooth scalar 
field like 
\begin{eqnarray}
V(x) := \xi R+m^2 +V'(x) \label{V}\:.
\end{eqnarray}
We also assume that $V'$ satisfies $\partial_t V'=0$ so that the space of solutions of (\ref{evolution}) is invariant under $t$-displacements.
Furthermore $\xi \in \R$, is a constant,
 $R$ is the scalar 
curvature and $m^2$ the squared mass of the particles associated to the field.
The domain of $P$ is the space of real-valued $C^\infty$ functions  compactly supported Cauchy data on $\Sigma$.
In the quasifree case, a straightforward way to define a QFT consistes of the assignment of a suitable
Green function of the operator $P$ \cite{fr}, 
in particular the Feynman propagator
$G_F(x,x')$ or, equivalently, the Wightman functions $W_{\pm}(x,x')$. Then the GNS theorem (e.g. see \cite{kw}) 
allows one to construct a corresponding Fock realization of the theory.
In a globally hyperbolic static spacetime it is possible to chose $t$-invariant  Green functions. In that case, in static coordinates,
one  performs the Wick rotation obtaining the Euclidean formulation of the same QFT.
This means that (locally) one can pass from the Lorentzian manifold
$(M,g)$ to a Riemannian manifold $(M_E, g^{(E)})$ by the analytic
continuation $t\to i\tau$ where $t,\tau \in \R$. This defines a (local) Killing
vector $\partial_\tau$ in the Riemannian manifold $M_E$ and a corresponding
(local) ``static'' coordinate frame $(\tau,\vec{x})$ therein. As is well-known 
\cite{fr},
when the orbits of
the Euclidean time $\tau$ are closed with period $\beta$, $T=1/\beta$ has to be interpreted as 
the temperature of the quantum state because the Wightman two-point function of the associated quasifree state satisfy the KMS condition at the inverse temperature $\beta$. 
In this approach, the Feynman propagator $G_F(t-t',\vec{x},\vec{x}')$
 determines -- and (generally speaking \cite{fr}) it is completely determined by --
 a proper Green function (in the spectral theory
sense) $S_\beta(\tau-\tau',\vec{x},\vec{x}')$
of a corresponding  self-adjoint extension $A$ of the operator
\begin{eqnarray}
A' :=  -\nabla^{(E)}_a\nabla^{(E) a} + V(\vec{x}) \::\:  C_0^\infty(M_E) \to L^2(M_E,d\mu_{g^{(E)}})\:.
\end{eqnarray}
$S_\beta(\tau-\tau',\vec{x},\vec{x}')$ is periodic with period $\beta$ in the $\tau-\tau'$ entry and it is
 said the Schwinger function. As a matter of fact, $S_\beta$ turns out to be the integral kernel of
$A^{-1}$ when $A>0$.

The {\em partition function}
 of the quantum state associated to $S_\beta$ is the functional
integral evaluated over the field configurations periodic with period $\beta$
in the Euclidean time
\begin{eqnarray}
Z_\beta = \int {\cal D}\phi \:e^{-S_E[\phi]}\:,
\end{eqnarray}
the Euclidean action $S_E$ being ($d\mu_{g^{(E)}} := \sqrt{g^{(E)}} d^4x$)
\begin{eqnarray}
S_E[\phi] = \frac{1}{2}\int_M d\mu_{g^{(E)}}(x) \:\phi(x) (A \phi)(x)\:.
\end{eqnarray}
Thus, extending the analogous result for finite dimensional Gaussian integral, one has 
\begin{eqnarray}
Z_\beta = \left\{ 
\det \left(\frac{A}{\mu^2}\right)\right\}^{-1/2}\:,\label{det}
\end{eqnarray}
where $\mu$ is a mass scale which is necessary for dimensional reasons. To give a sensitive interpretation of  that determinant,  the idea \cite{haw} is to try to exploit (\ref{det'}).\\
If $M_E$ is a $D$-dimensional Riemannian compact manifold 
and $A'$ is bounded below by some constant 
$b\geq 0$, $A'$ admits the {\em Friedrichs} self-adjoint extension $A$
which is also bounded below by the same bound of $A'$, moreover the spectrum
of $A$ is discrete and each eigenspace has a finite dimension.
Then, as we said,  one can consider the series 
\begin{eqnarray}
\zeta(s|A/\mu^2) := {\sum_j}' \left( \frac{\lambda_j}{\mu^2}\right)^{-s}\:.
\label{sum}
\end{eqnarray}
Remarkably \cite{haw,book}, in the given hypotheses, the series above converges for $Re$ $s >D/2$ and it is possible to continue the
 right-hand side above into a meromorphic function of $s$ which is regular
at $s=0$. Following  (\ref{det'}) and taking the presence of $\mu$ into account, we define:
\begin{eqnarray}
Z_\beta := e^{\frac{1}{2}\frac{d\:}{ds}|_{s=0}  \zeta(s|A/\mu^2)}\:, 
\label{zeta}
\end{eqnarray}
where  the function $\zeta$ on the right-hand side is the analytic continuation
of that defined in (\ref{sum}). 
It is possible the define the $\zeta$ function in terms of the heat kernel
of the operator $A$, $K(t,x,y|A)$ \cite{book}. This is the smooth integral kernel of
the (Hilbert-Schmidt, trace-class) operators
$e^{-tA}$, $t>0$. One has, for $Re$ $s>D/2$,
\begin{eqnarray}
\zeta(s|A/\mu^2) = \int_M d\mu_{g^{(E)}}(x) \int_0^{+\infty} 
dt\:\frac{\mu^{2s}t^{s-1}}{\Gamma(s)} \left[ K(t,x,x|A) -P_0(x,x|A) \right]
\:, \label{*}  
\end{eqnarray}
$P(x,y|A)$ is the integral kernel of the projector on the null-eigenvalues
eigenspace of $A$.\\
When $M_E$ is not compact, the spectrum of $A$ may included a continuous-spectrum part, however, 
it is still possible to generalize the 
definitions and the results above considering suitable integrals on the 
spectrum of $A$, provided $A$ is strictly positive
(e.g, see \cite{wald79}).\\
Another very useful tool is the {\em local} $\zeta$ function  
that can be  defined in two differen,t however equivalent, ways \cite{wald79,m1,book}:
\begin{eqnarray}
\zeta(s,x|A/\mu^2) =  \int_0^{+\infty} 
dt\:\frac{\mu^{2s}t^{s-1}}{\Gamma(s)} \left[ K(t,x,x|A) -P_0(x,x|A) \right]
\:, \label{**}  
\end{eqnarray}
and, $\phi_j$ being the smooth eigenvector of the eigenvalue $\lambda_j$,
\begin{eqnarray}
\zeta(s,x|A/\mu^2) = {\sum_j}' \left(\frac{\lambda_j}{\mu^2}\right)^{-s}
\phi_j(x)\phi^*_j(x)\:. \label{***}
\end{eqnarray}
Both the integral and the series converges for $Re$ $s>D/2$ and the local
zeta function  enjoys the same analyticity properties of the integrated
$\zeta$ function. For future convenience it is also 
useful to define, in the sense of the analytic continuation, 
\begin{eqnarray}
\zeta(s,x,y|A/\mu^2) =  \int_0^{+\infty} 
dt\:\frac{\mu^{2s}t^{s-1}}{\Gamma(s)} \left[ K(t,x,y|A) -P_0(x,y|A) \right]
\: \label{**'}  
\end{eqnarray}
(see \cite{m1,m2} for the properties of this off-diagonal $\zeta$-function).
In the 
framework of the $\zeta$-function regularization framework, 
the {\em effective Lagrangian} is defined as
\begin{eqnarray}
{\cal L}(x|A)_{\mu^2}
:= \frac{1}{2}\frac{d\:}{ds}|_{s=0}  \zeta(s,x|A/\mu^2)\:, 
\label{L}
\end{eqnarray}
and thus, in a thermal theory,
 $Z_\beta = e^{-S_\beta}$ where $S_\beta = \int d\mu_g 
{\cal L}_{\beta\mu^2}$.
A result which generalizes to any 
dimension an earlier results by Wald \cite{wald79} is the following \cite{m1}. The above-defined effective Lagrangian 
can be computed by a {\em point-splitting procedure}:
For $D$ even
\begin{eqnarray}
{\cal L}(y|A)_{\mu^2} &=& \lim_{x\to y}
\left\{ - \int_0^{+\infty}\frac{dt}{2t}K(t,x,y|A) - 
\frac{a_{D/2}(x,y)}{2(4\pi)^{D/2}}\ln\frac{\mu^2\sigma(x,y)}{2} \right.
\nonumber\\
&+& \left.  \sum_{j=0}^{D/2-1} (\frac{D}{2}-j-1)! \frac{a_j(x,y|A)}
{2(4\pi)^{D/2}} \left(\frac{2}{\sigma(x,y)} \right)^{D/2-j}\right\}
- 2\gamma \frac{a_{D/2}(y,y)}{2(4\pi)^{D/2}}\:,
\end{eqnarray}
for $D$ odd (notice that $\mu$ disappears)
\begin{eqnarray}
{\cal L}(y|A)_{\mu^2} &=& \lim_{x\to y}
\left\{ - \int_0^{+\infty}\frac{dt}{2t}K(t,x,y|A) - 
\sqrt{\frac{2}{\sigma(x,y)}}\frac{a_{(D-1)/2}(x,y)}{2(4\pi)^{D/2}} \right.
\nonumber\\
&+& \left.  \sum_{j=0}^{(D-3)/2} \frac{(D-2j-2)!!}{2^{(D+1)/2-j}} 
\frac{a_j(x,y|A)}
{2(4\pi)^{D/2}} \left(\frac{2}{\sigma(x,y)} \right)^{D/2-j}\right\}\:.
\end{eqnarray}
Above, $\sigma(x,y)$ is one half the square of the geodesical distance of
$x$ from $y$ and the coefficients $a_j$ are the well-known off-diagonal
coefficients of the small-$t$ expansion of the heat-kernel. These coefficients, in spite of 
their non symmetric definition, turns out to by invariant when interchanging $x$ and $y$ 
\cite{m3,m4}.

\section{Stress-energy tensor and field fluctuations}
\noindent {\bf 2.1} {\em Generalizations of the local $\zeta$ function 
technique.} Physically relevant quantities are the {\em (quantum) field fluctuation}
and the {\em averaged (quantum) stress tensor}, respectively:
\begin{eqnarray}
<\phi^2(x)> &=& \frac{\delta}{\delta J(x)}|_{J\equiv 0}
\ln \int {\cal D} \phi\: e^{-S_E + \int d\mu_{g^{(E)}} \phi^2 J} \label{phi}\:,\\
<T_{ab}(x)> &=& \frac{2}{\sqrt{g^{(E)}(x)}}\frac{\delta}{\delta g^{(E)ab}(x)}
\ln \int {\cal D} \phi\: e^{-S_E[g^{(E)}]} \:. \label{tab}
\end{eqnarray}
A standard method to compute them is the so-called {\em point-splitting procedure} \cite{bd,fu,waldlibro,m4,m5}.
 It is however possible to extend the $\zeta$-function technique \cite{m0,im,m1,m2,m4} to define suitable $\zeta$ functions
  regularizing  those 
quantities directly, 
similarly to what done for the effective Lagrangian.  Consider the stress tensor.
The idea relies upon
the following chain of formal identities \cite{m0}
\begin{eqnarray}
& &{\sqrt{g^{(E)}(x)}} <T_{ab}(x)> \mbox{``}=\mbox{''} 
2\frac{\delta}{\delta g^{(E) ab}(x)}
\ln Z_\beta \mbox{``}=\mbox{''}  \frac{\delta}{\delta g^{(E)ab}(x)}
\frac{d}{ds}|_{s=0} 
\zeta(s|A/\mu^2) \nonumber\\ 
& & \mbox{``}=\mbox{''} \frac{\delta}{\delta g^{(E)ab}(x)}\frac{d}{ds}|_{s=0} 
{\sum_j}'\left(\frac{\lambda_j}{\mu^2} \right)^{-s} \mbox{``}=\mbox{''} 
\frac{d}{ds}|_{s=0} \mu^{-2s}
{\sum_j}' \frac{\delta \lambda_j^{-s}}{\delta g^{(E)ab}(x)}\:. 
\end{eqnarray}
Following this route, one {\em define} the $\zeta$-regularized (or renormalized) stress tensor
as 
\begin{eqnarray}
<T_{ab}(x|A)>_{\mu^2}:=
\frac{1}{2} \frac{d}{ds}|_{s=0} Z_{ab}(s,x|A/\mu^2) \label{zetatab}\:,
\end{eqnarray}
where, {\em in the sense of the analytic continuation of the left-hand side}
\begin{eqnarray}
Z_{ab}(s,x|A/\mu^2) := 2{\sum_j}'\mu^{-2s}
 \frac{\delta \lambda_j^{-s}}{\delta g^{ab}(x)}\:. 
\label{zetab}
\end{eqnarray}
The difficult problem is now twofold: how to compute the functional derivative in the right-hand side of (\ref{zetab}) and
whether or not the series in the right-hand side of (\ref{zetab})
defines an analytic function of $s$ in a neighborhood of $s=0$.
We have the result \cite{m0,m2}:\\

\noindent{\bf Theorem 1.} 
{\em If $M_E$ is compact, $A\geq 0$ and $\mu^2>0$, then $Z_{ab}(s,x|A/\mu^2)$ is 
well-defined and is a $C^\infty$ function of $x$ which is also meromorphic 
in $s\in \C$. In particular, it is analytic in a neighborhood of $s=0$.}\\

\noindent The result above has been checked  even in several 
noncompact manifolds (containing singularities) \cite{m0,book}. In that case,
the summation in the right-hand side of (\ref{zetab}) has to be replaced by
a suitable spectral integration. The series in the right-hand side 
of (\ref{zetab}) can be explicitly computed as  \cite{m0,m2}:
\begin{eqnarray}
s{\sum_j}' \left\{ \frac{2}{\mu^2}\left(\frac{\lambda_j}{\mu^2}
\right)^{-s-1} 
T_{ab}[\phi_j,\phi^*_j](x)
+ g_{ab}(x) \left(\frac{\lambda_j}{\mu^2}\right)^{-s}\right\} \:,\nonumber
\end{eqnarray}
$T_{ab}[\phi_j,\phi^*_j](x)$ being the classical stress tensor
evaluated on the modes of $A$ (see \cite{m0,m2,book} for details). The series
converges for $Re$ $s>3D/2 +2$. \\
It is similarly possible to define a $\zeta$-function regularizing the field 
fluctuation \cite{im,m1}:
\begin{eqnarray}
<\phi^2(x|A)>_{\mu^2} := \frac{d}{ds}|_{s=0} \Phi(s,x|A/\mu^2)\:, \nonumber 
\end{eqnarray}
where 
\begin{eqnarray}
\Phi(s,x|A/\mu^2)    := \frac{s}{\mu^2} \zeta(s+1,x|A/\mu^2)
\:.
\end{eqnarray}
The properties of these functions have been studied in \cite{im,m1} and
several applications on concrete cases are considered ({\em e.g.}
cosmic-string spacetime and homogeneous spacetimes). In particular, in 
\cite{m1}, the problem of the change of the parameter $m^2$ in the field
fluctuations has been studied.\\

\noindent {\bf 2.2} {\em Physically meaningfulness of the procedures.} We are now interested in the physical meaningfulness of the presented
regularization techniques. The following general results strongly suggest that it is the case
\cite{m0,m2,m5}.\\

\noindent {\bf Theorem 2.}
{\em If $M_E$ is compact, $A\geq 0$ and $\mu^2>0$, and the averaged 
quantities above are those defined above in terms of local $\zeta$-function
regularization, then}

(a) {\em $<T_{ab}(x|A)>_{\mu^2}$ 
defines a $C^\infty$ symmetric tensorial field.} 

(b) {\em Similarly to the classical result,}
\begin{eqnarray} 
\nabla^b <T_{bc}(x|A)>_{\mu^2} = -\frac{1}{2} <\phi^2(x|A)>_{\mu^2}
 \nabla_c V'(x)\:.
\end{eqnarray}

(c)  {\em Concerning the trace of the stress tensor, it is naturally decomposed
in the classical and the known quantum anomalous part}
\begin{eqnarray}
g^{ab}(x)<T_{ab}(x|A) >_{\mu^2} &=& \left(\frac{\xi_D-\xi}{4\xi_D-1}\Delta 
-m^2 -V'(x) \right)<\phi^2(x|A)>_{\mu^2}\nonumber\\
&+& \delta_D \frac{a_{D/2}(x,x|A)}{(4\pi)^{D/2}}  - P_0(x,x|A)\:,
\end{eqnarray}
 {\em where $\delta_D =0$ if $D$ is 
odd and $\delta_D=1$ if $D$ is even, $\xi_D = (D-2)/[4(D-1)]$.} 

(d) {\em for any $\alpha>0 $}
\begin{eqnarray}
<T_{ab}(x|A)>_{\alpha\mu^2} = <T_{ab}(x|A)>_{\mu^2} + t_{ab}(x)\ln \alpha\:,
\end{eqnarray}
{\em where, the form of $t_{ab}(x)$ which depends on the geometry only and 
 is in agreement with Wald's axioms} \cite{waldlibro}, {\em has been given
in} \cite{m2,m5}.

(e) {\em In the case $\partial_0 = \partial_\tau$ is a global Killing vector,  
the manifold admits periodicity $\beta$ along the lines tangent to 
$\partial_0$ and $M$ remains smooth (near any fixed points of the
 Killing orbits) 
fixing arbitrarily $\beta$ in a 
neighborhood and, finally, $\Sigma$ is a global surface everywhere 
normal to $\partial_0$, then}
\begin{eqnarray}
\frac{\partial\:\:}{\partial \beta} \ln Z(\beta)_{\mu^2} =  
\int_\Sigma d\vec{x} \sqrt{g(\vec{x})}
 <T_0^0(x,\beta|A)>_{\mu^2}\:. 
\end{eqnarray}

\noindent Another general achievement regards the possibility to recover
the Lorentzian theory from the Euclidean one \cite{m2}:\\

\noindent{\bf Theorem 3.}
{\em Let  $M_E$ be compact, $A\geq 0$, $\mu^2>0$. Also assume that  $M_E$ is  
static with global Killing time $\partial_\tau$ and (orthogonal) global spatial
section $\Sigma$ and finally, $\partial_\tau V'\equiv 0$. Then} 

(a) $\partial_\tau <\phi^2(x|A)>_{\mu^2} \equiv 0$;

(b) $\partial_\tau {\cal L}(x|A)_{\mu^2} \equiv 0$;

(c) $\partial_\tau <T_{ab}(x|A)>_{\mu^2}   \equiv 0$;

(d) $ <T_{0i}(x|A)>_{\mu^2}   \equiv 0$ {\em for} $i=1,2,3,...,D-1$

\noindent {\em where
the averaged  quantities above are those defined above in terms of local 
$\zeta$-function regularization and coordinates $\tau \equiv x^0,
\vec{x}\in \Sigma$  are employed.}\\

\noindent These properties allow one to continue 
the Euclidean considered quantities  into imaginary values of 
the coordinate $\tau \mapsto it$ obtaining {\em real } functions of the
Lorentzian time $t$.\\
Some of the properties above (regarding Thm.1, Thm. 2, Thm.3) 
have been found to be valid  in some noncompact manifolds too (Rindler spacetime, cosmic string spacetime, 
Einstein's open spacetime, $H^N$ spaces, G\"{o}del spacetime, BTZ spacetime)
\cite{m0,im,caldarelli,radu0,radu1,radu2,radu3,bmvz,rf02,ss04,amr05}. In particular,
 the presented theory  has been 
successfully exploited to compute the quantum back reaction on the three-dimensional
 BTZ metric  \cite{bmvz} in the case of the singular ground state containing
 a naked singularity. A semiclassical implementation of the cosmic
 censorship conjecture has been found in that case.\\

\noindent {\bf 2.3.} {\em Interplay of zeta-function approach and  point-splitting technique.}
The procedure of the point-splitting to renormalize the field 
fluctuation as well as the  stress tensor  \cite{bd,waldlibro,m4,m5}, when the two-point functions are referred 
to {\em quasifree Hadamard-states},
 can be summarized as
\begin{eqnarray}
<\phi^2(y)>_{\scriptsize \mbox{ps}} 
&=& \lim_{x\to y} \left\{ G(x,y) - H(x,y) \right\} 
\label{ps1}\:,\\
<T_{ab}(y)>_{\scriptsize \mbox{ps}} &=& \lim_{x\to y} {\cal D}_{ab}(x,y)
\left\{ G(x,y) - H(x,y) \right\} 
+ g_{ab}(y) Q(y)\:,\label{ps2}
\end{eqnarray}
where $G(x,y)$ is the symmtric part the two-point Wightman function of the considered quantum state
or, in Euclidean approach, the corresponding Schwinger function.
$H(x,y)$ is the {\em Hadamard parametrix} which depends on the {\em local} 
geometry only and takes the short-distance singularity into account. 
$H(x,y)$ is represented in terms of a truncated series of functions of 
$\sigma(x,y)$.
The operator ${\cal D}_{ab}(x,y)$ is a bi-tensorial operator obtained by
``splitting'' the argument of the classical expression of the stress tensor
(see \cite{m2,m5}). Finally
$Q(y)$ is a scalar obtained by imposing several physical 
conditions (essentially, the appearance of the conformal anomaly,
the conservation of the stress tensor and the
 triviality of the Minkowskian limit)
\cite{waldlibro} in the
 left-hand side of (\ref{ps2}) (see 
\cite{bd,fu,waldlibro,m2} for details). More recently, in the framework of Lorentzian generally locally covariant algebraic quantum field theory 
in curved spacetime,
it has  established \cite{m5}
that $Q$ can be omitted, redefining the classical stress-energy tensor, and thus ${\cal D}_{ab}(x,y)$, into a way that it does not affect the classical expression of 
$T_{\mu\nu}$ when computed on solutions of the equations of motion,
 improving the point-splitting procedure. See \cite{hack} where that point-splitting procedure is discussed and applied especially to cosmology.  
In geodesically convex neighborhoods:
\begin{eqnarray}
H_\mu(x,y) &=& \frac{\sum_{j=0}^L u_j(x,y) \sigma(x,y)^j}{(4\pi)^{D/2}
(\sigma(x,y)/2)^{D/2-1}} + \delta_D \left[\sum_{j=0}^{M} v_j(x,y) 
\sigma(x,y)^j \ln \left(\frac{\mu^2\sigma(x,y)}{2}\right)\right] \nonumber\\
&+& \delta_D \sum_{j=0}^{N} w_j(x,y) \sigma(x,y)^j \label{hadamard}\:.
\end{eqnarray}
$L,M,N$ are fixed integers (see \cite{m2,m5} for details), 
$\delta_D=0$ if $D$ is odd and $\delta_D=1$ otherwise. The coefficients
$u_j$ and $v_j$ are smooth functions of $(x,y)$ which are completely 
determined by the local 
geometry. The coefficients $w_j$ are determined
once one has fixed $w_0$, and they are omitted \cite{m5} when dropping $Q$. Dealing with Euclidean approaches, it is possible 
to explicitly compute $u_j$ and $v_j$ in terms of heat-kernel coefficients \cite{m1,m2}. One has the following result \cite{m1,m2}.\\

\noindent {\bf Theorem 4.}
{\em If $M_E$ is compact, $A\geq 0$ and $\mu^2>0$, and the averaged 
quantities in the left-hand side below are those defined above in terms of local $\zeta$-function
regularization, then}
\begin{eqnarray}
<\phi^2(y|A)>_{\mu^2} &=& \lim_{x\to y} \left\{ G(x,y) - H_{\mu'}(x,y) \right\} 
\label{ps1*}\:,\\
<T_{ab}(y|A)>_{\mu^2} &=& \lim_{x\to y} {\cal D}_{ab}(x,y)
\left\{ G(x,y) - H_{\mu'}(x,y) \right\} + g_{ab}(y) Q(y)
\label{ps2*}\:,
\end{eqnarray}
{\em where $G(x,y) = \zeta(1,x,y|A/\mu^2)$ given in} (\ref{**'}),
{\em $H_{\mu'}$ is completely determined by (\ref{hadamard}) with the requirement}
\begin{eqnarray}
w_0(x,y) := -\frac{a_{D/2-1}(x,y|A)}{(4\pi)^{D/2}}[2\gamma + \ln {\mu'}^2]\:,
\end{eqnarray}
{\em and the term $Q$ is found to be}
\begin{eqnarray}
Q(y) = \frac{1}{D}\left(-P_0(y,y|A) +\delta_D \frac{a_{D/2}(y,y|A)}{(4\pi)^{D/2}}\right) \:.
\end{eqnarray}
{\em If ${\cal D}_{ab}$ is defined in order to drop $Q$ in the right-hand side  of (\ref{ps2*}),  $H_{\mu^2}$ is determined by fixing $w_0(x,y)=0$.
The scales $\mu$ and $\mu'$ satisfies $\mu=c\mu'$ for some constant $c>0$.}

\end{document}